\documentclass[12pt]{article}
\usepackage{amsmath}
\usepackage{amssymb}
\usepackage{amsfonts}
\newcommand{\be}{\begin{equation}}
\newcommand{\ee}{\end{equation}}
\newcommand{\beq}{\begin{equation}}
\newcommand{\eeq}{\end{equation}}
\newcommand{\bea}{\begin{eqnarray}}
\newcommand{\eea}{\end{eqnarray}}
\newcommand{\vep}{\varepsilon}
\usepackage{color}

\usepackage{epsfig}
\usepackage{graphicx}
\usepackage{hyperref}
\newcommand{\email}[1]{\footnote{#1}}
\begin{document}
\baselineskip=22pt
%
\begin{flushright}
\hfill{\bf BIHEP-TH-2004-06}\\
\hfill{\bf USTC-ICTS-04-03}\\
\end{flushright}
%
%
\begin{center}
{\Large \bf Statistical properties of radiation fields\\
in a compact space} \vspace*{1cm}

Zhe Chang\email{changz@mail.ihep.ac.cn}\\
{\em Institute of High Energy Physics, Chinese Academy of Sciences}\\
{\em P.O.Box 918(4), 100039 Beijing, China}\\
Cheng-Bo Guan\email{guancb@ustc.edu.cn}\\
{\em Interdisciplinary Center for Theoretical Study}\\
{\em University of Science and Technology of China, 230026 Hefei, China}

\end{center}
\vspace*{2.0cm}

%
\begin{abstract}

We discuss radiation fields in a compact space of finite
size instead of that in a cavity for investigating the coupled
atom-radiation field system. Representations  of $T(1)\times
SO(4)$ group are used to give a formulation for kinematics of the
radiation fields.  The explicit geometrical
parameter dependence of statistical properties of radiation
fields is obtained. Results show  remarkable differences from that
of the black-body radiation system in free space. \vspace*{1.0cm}

\begin{flushleft}
\end{flushleft}
\end{abstract}

\newpage

%
\section{Introduction}

The development of a single-atom maser allows a detailed study of
the atom-field interaction\cite{maser1}-\cite{maser4}. The
realization of a single-atom maser has been made possible due to
the enormous progress in the construction of superconducting
cavities together with the laser preparation of highly excited
atoms called Rydberg atoms. Resonant effects\cite{reso} associated
to the coupling of atoms with strong radiofrequency field has been
observed. In particular, it has been demonstrated that the
spontaneous emission rate of an atom inside the cavity is
different from its value in free space\cite{spont1}-\cite{spont5}.
This effect can be discussed from several different approaches,
e.g., attributed it to a change of the spectral density of the
modes of the vacuum radiation field due to the cavity's resonating
structure. The theoretical understanding of these effects by
making use of perturbation theory requires the calculation of very
high-order terms. It makes the standard Feynman diagram technique
practically unreliable in the case. On the other side, the
nonlinear character of the problem involved in realistic
situations can not be ignored simply. A naive solution of this
difficulty may be to assume that under certain conditions, the
coupled atom-radiation field system can be approximated by a system
composed of a harmonic oscillator coupled linearly to the
radiation field through some effective coupling constants. Thus, a
significant number of works have been sparked to the study of
cavity QED.

Another motivation to study the radiation field system in cavity comes
from the cosmology.  Astronomical observations\cite{CMBR}-\cite{WMAP}
have provided plenty of
supports to the cosmological principle\cite{RW,SW}, which states
that the universe is spatially homogeneous and isotropic on large scales.
This principle implies that we can build up a comoving coordinate system,
in which the spatial part has a maximal symmetry. In general,
there are three types of global symmetry for the three-dimensional
maximally symmetric space, {\em i.e.}, $ISO(3)$, $SO(4)$ and
$SO(1,3)$. Usually, they are called the flat, closed and open
universe, respectively. In order to understand the evolution of
the universe, the whole distribution and activity of matters
(including radiations) in space have to be taken into account. And
we would like to point out that, if properties of matters in space
as a whole are to be studied, the difference of global symmetries should
not be neglected. In particular, we know that the early epoch of the
universe was governed by the radiation fields. Therefore, a careful
study of radiation fields in a compact space is necessary.

In free space, it is well-known that the radiation fields can be
described by an ideal gas model (a thermodynamic system consisting
of free massless particles). However, one can not extend this idea
to the case of compact space straightforwardly.  In order to
investigate the quantum statistical properties of radiation fields
in a compact space, we begin with constructing of the
representation of one-particle states. We notice that the
space-time $\mathbb{R}^{1}\times S^3$ is a conformally deformed
de Sitter space-time. This space-time has a translation symmetry
in the time direction and six rotation symmetries in the spatial part.
That is to say, the isometry group of the space-time $\mathbb{R}^{1}\times S^3$ is
$T(1)\times SO(4)$. By making use of the representation theory of
this group, we get a representation of one-particle states as well
as the corresponding dispersion relation for the radiation field.
The density of state with respect to the energy spectrum is
continuous in the case of the compact space with a large radius.
It should be pointed out that the obtained density of state is
different from that of a black body in free space. The
geometrical parameter dependence of statistical properties of
the radiation field system in a compact space can be obtained explicitly.
It has been a goal of cavity QED for a long time.

This paper is organized as follows. In Section 2, we will discuss
the kinematics of the radiation fields in the compact space $S^3$.
The Hilbert space of free photons as well as the corresponding
dispersion relations is presented. The Bose-Einstein statistics of
the radiation fields is intensively studied in Section 3.
Statistical properties of the radiation fields, which is remarkably
different from that in free space, are shown. We then give the conclusions
and remarks in section 4.

\section{The kinematics of radiation fields}

The metric of the compact space-time $\mathbb{R}^{1}\times S^3$ is
of the form
\begin{equation}\label{metric}
ds^2 = dt^2 - R^2d\Omega^2_3~,
\end{equation}
where $R$ denotes the radius of the sphere $S^3$. The spatial part of the
metric,  $d\Omega^2_3$, can be expanded with Euler angular variables
\begin{equation}\label{sphere1}
d\Omega^2_3 = dr^2 + \sin^2r(d\theta^2 + \sin^2{\theta}d\phi^2)~,
\end{equation}
or with a unit vector in the four-dimensional Euclidean space
\begin{eqnarray}\label{sphere2}
&&d\Omega^2_3 = dx^idx^i~,\\
&&\sum_{i}x^ix^i = 1~,~~~(i=1,2,3,4)~.\nonumber
\end{eqnarray}
The isometry group of the space-time is $T(1)\times SO(4)$.
Infinitesimal elements of the isometry group can be expressed in
terms of the following generators
\begin{equation}\label{operator}
\hat{H}=-i\partial_{t}~,~~~ \hat{M}^{ij}= -i(x^i\partial_j-x^j\partial_i)~.
\end{equation}
Here $\hat{H}$ denotes the Hamiltonian and $\hat{M}^{ij}$ are
angular momentums.
\newline
The motion of a free photon is described by the Klein-Gordon
equation
\begin{equation}\label{motion}
\Box A^{\lambda}= \frac{1}{\sqrt{-g}}\partial_{\mu}\left(\sqrt{-g}
g^{\mu\nu}\partial_{\nu}\right)A^{\lambda} = 0~.
\end{equation}
Here the d'Alembertian operator $\Box$ is an invariant of $T(1)\times
SO(4)$
\begin{equation}\label{casimir}
\Box = -(\hat{H}^2-\frac{1}{2}\hat{M}^{ij}\hat{M}_{ij}R^{-2})~~.
\end{equation}
It is convenient to introduce the set of operators
$\hat{L}_\alpha$ and $\hat{P}_\alpha$ as\cite{Thomas-1941}
\begin{eqnarray}\label{intro-LP}
&&\hat{L}_{\alpha}=\frac{1}{2}\epsilon_{\alpha\beta\gamma}\hat{M}^{\beta\gamma}~,\nonumber\\
&&\hat{P}_{\alpha}=\hat{M}^{4\alpha}~,~~(\alpha,~\beta,~\gamma~=1,2,3)~,
\end{eqnarray}
and their linear combinations,
\begin{eqnarray}\label{intro-JS}
&&\hat{J}_{\alpha}=\frac{1}{2}(\hat{L}_{\alpha}+\hat{P}_{\alpha})~,\nonumber\\
&&\hat{S}_{\alpha}=\frac{1}{2}(\hat{L}_{\alpha}-\hat{P}_{\alpha})~.
\end{eqnarray}
Then, one can separate the $so(4)$ Lie algebra $\{\hat{M}^{ij}\}$ into
two parts, which commutates with each other
\begin{eqnarray}\label{separation}
&&[\hat{J}_{\alpha},~\hat{S}_{\beta}] = 0~,\nonumber\\
&&[\hat{J}_{\alpha},~\hat{J}_{\beta}] = i\epsilon_{\alpha\beta\gamma}\hat{J}_{\gamma}~,\nonumber\\
&&[\hat{S}_{\alpha},~\hat{S}_{\beta}] = i\epsilon_{\alpha\beta\gamma}\hat{S}_{\gamma}~,\nonumber\\
&&\frac{1}{2}\hat{M}^{ij}\hat{M}_{ij} = \hat{P}^2+\hat{L}^2= 2(\hat{J}^2+\hat{S}^2)~.
\end{eqnarray}
In other words, the $SO(4)$ group can be  expressed locally as a
direct product of two $SO(3)$ groups. Nevertheless, it is easy to
see that,
\begin{eqnarray}\label{lp0}
&&\hat{\bf L}\cdot\hat{\bf P} = 0~,\nonumber\\
&&\hat{J}^2 = \hat{S}^2~.
\end{eqnarray}
And then, we obtain the representation of the $so(4)$ universal Lie algebra $(\ref{intro-JS})$
\begin{eqnarray}\label{rep1}
&&\hat{J}^2 |j,j_3,s_3\rangle = j(j+1)|j,j_3,s_3\rangle~,~~
  \hat{J}_3 |j,j_3,s_3\rangle = j_3|j,j_3,s_3\rangle~,\nonumber\\
&&\hat{S}^2 |j,j_3,s_3\rangle = j(j+1)|j,j_3,s_3\rangle~,~~
  \hat{S}_3 |j,j_3,s_3\rangle = s_3|j,j_3,s_3\rangle~,\\
&&\frac{1}{2}\hat{M}^{ij}\hat{M}_{ij}|j,j_3,s_3\rangle = 4j(j+1)|j,j_3,s_3\rangle~,\nonumber
\end{eqnarray}
where $j=0,1,2,\cdots$ and $j_3=-j,\cdots,0,\cdots,j;~s_3=-j,\cdots,0,\cdots,j$~.
\newline
The representation of the translation group $T(1)$ is of the form
\begin{equation}\label{rep2}
\hat{H}|\varepsilon\rangle = \varepsilon|\varepsilon\rangle~.
\end{equation}
Now we can write the wave-function $A^{\lambda}$ into the form
\begin{equation}\label{rep3}
|\varepsilon;j,j_3,s_3;\sigma\rangle~,
\end{equation}
where $\sigma=\pm 1$ denotes the degrees of freedom of spin. The
corresponding dispersion relation for this eigen-state is given by
Eqs. $(\ref{motion})$ and $(\ref{casimir})$
\begin{equation}\label{dispersion}
\vep^2=4j(j+1)R^{-2}~.
\end{equation}
The Hilbert space of a free photon can therefore be parameterized as
\begin{equation}\label{states}
|j,j_3,s_3;\sigma\rangle~.
\end{equation}

\section{Statistical properties of radiation fields}

To conveniently study statistical properties of the radiation fields in a
compact space, one should compute the density of state for energy
spectrum first. The dispersion relation (\ref{dispersion}) can be
rewritten into the following form
\begin{equation}\label{dispersion1}
\vep^2 = \left[(2j+1)^2-1\right]R^{-2}=K^2-R^{-2}~,
\end{equation}
where we have introduced $K\equiv (2j+1)R^{-1}~$ for convenience.

For a given energy level $\vep$,  the value of $K$ (or $j$) is
determined. There exist $2K^2R^2$ eigen-states at this energy
level resulted by summing over the index parameters $(j_3,s_3)$
and $\sigma$. If the scale of the radius $R$ is large enough, one
can regard the spectrum of energy and $K$ as a continuous
distribution. Then, we get a measure of states with respect to the energy
spectrum, which is just the density of state,
\begin{eqnarray}\label{density}
&&dN_{\vep}= 2R^3K^2dK= 2R^3(\vep^2+R^{-2})^{1/2}\vep d\vep~,\nonumber\\
&&\rho(\vep)=2R^3(\vep^2+R^{-2})^{1/2}\vep ~.
\end{eqnarray}
The density of state is obviously different from that of the
ordinary black-body radiation system in free space, which is proportional
to the square of energy ($\vep^2$).

The photons obey the Bose-Einstein statistics.  The mean number of
photons on an eigen-state with energy $\vep$ is\cite{Stats}
$$
\displaystyle \frac{1}{e^{\vep/T}-1}~,
$$
where $T$ is the temperature of this system\footnote{For
simplicity, we have taken the chemical potential $\mu$ to be zero
and hence $F=\Omega+\mu N=\Omega$.}.
\newline
Immediately, we obtain a modified Planck distribution for the
radiation fields in the compact space $S^3$,
\begin{equation}\label{Planck}
\frac{dU}{V}= \displaystyle \frac{1}{V}\frac{\rho(\vep)\vep}{e^{\vep/T}-1}d\vep
  = \frac{1}{\pi^2}\frac{(\vep^2+R^{-2})^{1/2}\vep^2
  }{e^{\vep/T}-1}d\vep~.
\end{equation}
The distribution function represents the amount of radiation
energy in the spectral interval $\vep\sim(\vep+d\vep)$ and
a unit volume. In picture {\bf 1}, we give a sketch for this
modified Planck distribution comparing with the ordinary one in
free space.

\begin{figure}[htb]\label{pic}
\begin{center}
\includegraphics[height=70mm,width=100mm]{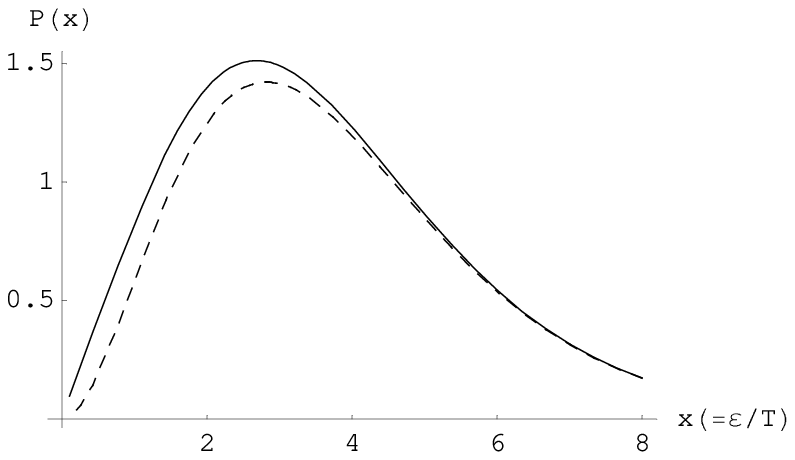}
\caption{\scriptsize Planck distribution of radiation fields on
$S^3$ (real line, $RT=1$) and the ordinary distribution of a
black-body radiation system in free space (dashed line).}
\end{center}
\end{figure}
The total number of photons of the radiation system can be
expressed as
\begin{equation}\label{total-number}
N= \displaystyle\int_{0}^{\infty}\frac{\rho(\vep)}{e^{\vep/T}-1}d\vep
 = \displaystyle
 2R^3T^3\int_{0}^{\infty}\frac{(x^2+(RT)^{-2})^{1/2}x}{e^x-1}dx~.
\end{equation}
When the parameter $RT$ runs to infinity, we recover
the familiar result  of the black-body radiation system in free space. In
the follows, we will calculate quantities by expanding them in terms of
$(RT)^{-1}$ to see clearly the deviation from those got in free
space.

The thermodynamic quantity of total energy is
\begin{equation}\label{total-energy}
U= \displaystyle\int_{0}^{\infty}\frac{\rho(\vep)\vep}{e^{\vep/T}-1}d\vep
 \simeq \displaystyle 12\zeta(4)R^3T^4+ \zeta(2)RT^2~,
\end{equation}
where $\zeta(m)$ ($\equiv\sum^{\infty}_{n=1}1/n^m$) is the Riemann
zeta-function.
\newline
The free energy and thermodynamic potential can also be calculated
easily,
\begin{eqnarray}\label{total-potential}
&F= \Omega&= -PV= \displaystyle T\int_{0}^{\infty}\rho(\vep)\log(1-e^{-\vep/T})d\vep\\
&&\simeq -\displaystyle\left(4\zeta(4)R^3T^4 + \zeta(2)RT^2\right)~.\nonumber
\end{eqnarray}
We know that the volume of the three-dimensional sphere with radius
$R$ is $2\pi^2R^3$. Hence the radiation pressure is
\begin{equation}\label{total-pressure}
P= \displaystyle -\frac{F}{V}\simeq
   \frac{1}{2\pi^2}\left(4\zeta(4)T^4+\zeta(2)R^{-2}T^2\right)~.
\end{equation}
Furthermore, by making use of the relations among thermodynamic
quantities
\begin{equation}\label{total-relation}
U= F+TS~~~{\rm or}~~~dF=-SdT-PdV~,
\end{equation}
we can obtain the entropy of radiation fields
\begin{equation}\label{total-entropy}
S\simeq 16\zeta(4)R^3T^3+ 2\zeta(2)RT~.
\end{equation}

In an adiabatic process experienced by the radiation fields,
the entropy doesn't change.  So the
product $RT$ can be viewed as a constant in the
process. The total number of photons is also kept invariant in the
adiabatic process. But the other thermodynamic quantities will
vary with the geometrical parameter $R$ as follows,
\begin{equation}\label{adiabatic}
U,~F,~\Omega~\propto R^{-1}~,~~~P~\propto R^{-4}~.
\end{equation}
To discuss the equation of state of this radiation system,
we now calculate the density of thermal energy,
\begin{equation}\label{density-energy}
\rho = \frac{U}{V} \simeq \frac{1}{2\pi^2}\left(12\zeta(4)T^4+ \zeta(2)R^{-2}T^2\right)~.
\end{equation}
The equation of state of the radiation fields in the compact space
can be read out from Eqs. (\ref{total-pressure}) and
(\ref{density-energy}),
\begin{equation}
\rho\simeq 3P-\frac{\zeta(2)}{\pi\sqrt{2\zeta(4)}R^2}P^{1/2}~.
\end{equation}
It is obvious that, in the case of a large radius of the compact
space, the ordinary equation of state for the radiation fields in free
space can be recovered.

%
%
\section{Conclusions}

A good understanding of the resonant effect and spontaneous
emission rate difference in cavity requires first to know the full
space dependence of the cavity-induced damping and level shifts.
However, we know that it is a really difficult mathematical task.
Thus, many useful approximation methods have been developed.  As a
first step on the way of getting an exact solution of the problem,
we suggested working on a compact space of finite size instead of
a cavity. By making use of the representation theory of the
$T(1)\times SO(4)$ group, we studied the kinematics for a free
photon propagating in the  space-time ${\mathbb{R}}^1\times S^3$.
The explicit geometrical parameter dependence of statistical
properties of the radiation fields was presented. Results showed
remarkable differences from that of the black-body radiation in
free space.

The resonant effects associated to the coupling of atoms with
strong radiofrequency field as well as the spontaneous emission
rate can also be computed in a straightforward way. It is expected
that the nonlinear character of the coupled atom-radiation field
may be attributed to the geometrical parameters of the compact
space. These results will be published in a forthcoming paper.

In fact, the statistical properties of radiation fields in a
compact space are crucial to the evolution of the early universe.
In the early universe, radiation field is the dominant matter. A bit
difference of statistical properties of the radiation system can
generate observable effects on our present cosmology. All of these
studies are in progressing.

\vspace*{1.0cm}
\noindent
{\large\bf Acknowledgement:}\\
One of us (C.B.G) would like to thank Dr.\ Yu-Qi Li and
Xiao-Jun Wang for useful discussion and  suggestions. The work
was supported partly by the Natural Science Foundation of China.
C. B. Guan is supported by grants through the ICTS (USTC) from
the Chinese Academy of Sciences.

%

%
%

\begin{thebibliography}{999}
\bibitem{maser1}{D. Meschede, H. Walther, and G. M\"{u}ller, Phys. Rev. Lett. {\bf 54}, 551 (1985).}
\bibitem{maser2}{G. Rempe, H. Walther, and N. Klein, Phys. Rev. Lett. {\bf 58}, 353 (1987).}
\bibitem{maser3}{P. Filipowicz, J. Javanainen, and P. Meystre, Phys. Rev. A{\bf 34}, 3077 (1986).}
\bibitem{maser4}{L. A. Lugiato, M. O. Scully, and H. Walther, Phys. Rev. A{\bf 36}, 740 (1987).}
\bibitem{reso}{J. M. Winter, Ann. Phys. (Paris) {\bf 4}, 745 (1959).}
\bibitem{spont1}{K. H. Drexhage, in {\em Progress in optics XII},
                 edited by E. Wolf (North-Holland Publishing Comp., Amsterdam, 1974), p.165.}
\bibitem{spont2}{D. Kleppner, Phys. Rev. Lett. {\bf 47}, 233 (1981).}
\bibitem{spont3}{P. Goy, J. M. Raimond, M. Gross, and S. Haroche, Phys. Rev. Lett. {\bf 50}, 1903 (1983).}
\bibitem{spont4}{Yifu Zhu, A. Lezama, T. W. Mossberg, and M. Lewenstein, Phys. Rev. Lett. {\bf 61}, 1946 (1988).}
\bibitem{spont5}{S. E. Morin, C. C. Yu, and T. W. Mossberg, Phys. Rev. Lett. {\bf 73}, 1489 (1994).}
%
\bibitem{CMBR}{A. A. Penzias and R. W. Wilson, Astrophys. J. {\bf 142}, 419 (1965).}
\bibitem{COBE}{C. L. Bennett {\it et al}., Astrophys. J. Lett. {\bf 464}, L1 (1996).}
\bibitem{WMAP}{C. L. Bennett {\it et al}., Astrophys. J. Suppl. {\bf 148}, 1 (2003).}
\bibitem{RW}  {A. Friedmann, Z. Phys., {\bf 10}, 377 (1922); Z. Phys., {\bf 21}, 326 (1924);
               H. P. Robertson, Astrophys. J., {\bf 82}, 284 (1935);
                                Astrophys. J., {\bf 83}, 187, 257 (1936);
               A. G. Walker, Proc. Lond. Math. Soc. (2), {\bf 42}, 90 (1936).}
\bibitem{SW}  {S. Weinberg, {\it Gravitation and Cosmology: Principles and Applications
               of the General Theory of Relativity} (John Wiley \& Sons, New York, 1972).}
%
\bibitem{Thomas-1941}{L. H. Thomas, Ann. of Math. {\bf 42}, 113 (1941).}
\bibitem{Stats} {L. D. Landau and E. M. Lifshitz, {\it Statistical Physics}, Part 1,
                 3rd edition, trans. by J. B. Sykes and M. J. Kearsley (Butterworth-Heinemann, reprinted
                 by Beijing World Publishing Corp., 1999).}
%
%
\end{thebibliography}
\end{document}